\def\be{\begin{equation}}
\def\ee{\end{equation}}
\def\ber{\begin{eqnarray}}
\def\eer{\end{eqnarray}}
\def\bers{\begin{eqnarray*}}
\def\eers{\end{eqnarray*}}

\def\JPCM{J. Phys.: Condens. Matter}
\def\PR{{ Phys. Rev.}\ }
\def\PRL{{ Phys. Rev. Lett.}\ }

\def\JPCM{{ J. Phys.: Condens. Matter}\  }

\documentclass[aps,prb,twocolumn,showpacs,groupedaddressamsmath,amssymb]{revtex4-1}
\usepackage{dcolumn}   
\usepackage{amsmath}
\usepackage{graphicx}    
\usepackage{subfigure}
\usepackage{color}
\newcommand{\comment}[1]{}
\usepackage{ifthen}
\newboolean{includefigs}
\setboolean{includefigs}{true}      
\newboolean{includetext}
\setboolean{includetext}{true}     
\newcommand{\condcomment}[2]{\ifthenelse{#1}{#2}{}}
\begin{document}
\title[Interplay of force constants in the lattice dynamics of disordered alloys : An ab-initio study]{Interplay of force constants in the lattice dynamics of disordered alloys : \\
An ab-initio study}
\author{Rajiv K. Chouhan,$^{1,2}$ Aftab Alam,$^{3\dagger}$, Subhradip Ghosh,$^{4}$ and Abhijit Mookerjee$^{1}$}
\email{$^\dagger$Corresponding author: aftab@phy.iitb.ac.in; rajivchouhan@gmail.com}
\affiliation{$^{1}$Department of Condensed Matter Physics and Materials Science, S.N. Bose National Center for Basic Sciences, Kolkata 700098, India }
\affiliation{$^{2}$Theoretical Sciences Unit, Jawaharlal Nehru Centre for Advanced Scientific Research, Jakkur, Bangalore 560064, India }
\affiliation{$^{3}$Department of Physics, Indian Institute of Technology, Bombay, Powai, Mumbai 400076, India}
\affiliation{$^{4}$Department of Physics, Indian  Institute of Technology, Guwahati,  Assam, India}

\begin{abstract}
A reliable prediction of interatomic force constants in disordered alloys is 
an outstanding problem. This is due to the need for a proper treatment of 
multisite (at least pair) correlation within a random environment. The situation 
becomes even more challenging for systems with large difference in atomic size
and mass. We propose a systematic density functional theory (DFT) based study 
to predict the ab-initio force constants in random alloys. The method is based
on a marriage between special quasirandom structures (SQS) and the augmented
space recursion (ASR) to calculate phonon spectra, density of states (DOS) etc.
bcc TaW and fcc NiPt alloys are considered as the two distinct test cases. 
Ta-Ta (W-W) bond distance in the alloy is predicted to be smaller (larger) than
those in pure Ta (W), which, in turn, yields stiffer (softer) force constants
for Ta (W). Pt-Pt force constants in the alloy, however, are predicted to be 
softer compared to Ni-Ni, due to large bond distance of the former. Our
calculated force constants, phonon spectra and DOS are compared with experiments
and other theoretical results, wherever available. Correct trend of present 
results for the two alloys pave a path for future studies in more complex
alloy systems.
\end{abstract} 
\date{\today}
\maketitle
\section{Introduction}
{\par}In spite of years of rigorous research, a reliable \textit{ab-initio} theoretical model
for structural and/or substitutional disordered alloys is still lacking. More 
importantly, the understanding of the problem of lattice dynamics (or phonons)
in disordered alloys is still in its adolescence. This is mainly due to the 
presence of off-diagonal (multisite) disorder arising out of the force constant
tensor $D_{\mu\nu} (R_i-R_j)$ in the phonon problem. In addition the sum 
rule $D_{\mu\nu}(R_i)=-\sum_{R_j}D_{\mu\nu} (R_i-R_j)$ 
 makes the disorder at a site depend upon its neighborhood or the so called environmental disorder.
As such first principles method based on e.g. coherent potential approximation
(CPA)\cite{CPA} is inapplicable in this problem. Within CPA, the atoms
occupancy are assumed in an average sense embedded in a structure-less 
uniform average medium. This prohibits CPA to include structural relaxations, 
which contrasts from the experimental observations because bond distance
between atomic pairs (e.g. A-A, B-B, and A-B in a binary alloy) are generally
different.\cite{bond_dist} Various generalizations\cite{general_CPA1,general_CPA2} of CPA
have been suggested over time, each has its own advantages and
disadvantages. An striking approach, emerged in recent years, is the so called
special quasirandom structure (SQS) proposed by Zunger {\emph {et al.}}
\cite{Zunger90}, which carries the signature of configuration correlation with 
them. In particular,  SQS is an ordered supercell which is constructed in such 
a way to mimic the most relevant pair and multisite correlation functions of the 
disordered phase. Unlike CPA and other related approaches, SQS is a local
structural model which captures the most relevant microstates of disordered 
phase.

{\par} As far as the calculation of force constants for random alloys are
concerned, three approaches are mainly utilized in the past. The first 
attempt\cite{general_CPA2,Alam04} was to fit an empirical set of force constants
to match the available experimental phonon spectra. The second approach was
to compute the force constants from selected ordered structures and then use
them for random alloys.\cite{Alam07} This is of course not a proper solution,
because dynamical matrices are not directly transferable across the environment.
\cite{Walle02} In later studies\cite{Gan06}, few SQS methods have 
been used, but only $\le 8$-atoms cell were used which is not enough to capture
the detailed properties of phonons, as illustrated in the present work. 

{\par} In addition to a reliable estimate of the force constants, equally 
important is the need for a method to perform configuration average over the disorder
environment to get the phonon spectra, DOS etc. for the alloys. 
Augmented space formalism (ASF)\cite{Alam04} is a powerful
technique to do the same, and has been described in great details in many of
our earlier papers. Interested readers are referred to articles
\onlinecite{Alam04,Alam07} for any details related to formalism.
In the present rapid communication, we integrate a firs principles SQS method with the ASF to 
demonstrate the interplay of force constants within a disorder environment.
Unlike previous approaches, a systematic calculation of the force constants
with increasing size of the SQS cell is made. Stress on the atomic sites are
directly related to the force constant matrix and hence a small disturbance leads
to a large change. To overcome this effect, we use the SQS cell in conjunction 
with the {\it small displacement method}\cite{alfe} to construct the dynamical
matrix $D_{\mu\nu}$. Based on the predicted bond length distribution and the 
calculated force constants for each pairs A-A, B-B, and A-B, it is concluded
that a minimum of $32$-atom SQS cell is needed to capture the important 
disorder correlations, and hence a reliable phonon dispersion. Two different
alloy systems, bcc TaW and fcc NiPt, are chosen to demonstrate the reliability 
of the approach. Both the systems have inelastic neutron scattering data to 
compare our theoretical results. Ta and W belong to $5$d-metal series with
similar size and atomic masses, but quite different force constants. Ni and Pt
on the other hand differ significantly in size ($\sim~ 12\%$), masses (Pt is
$3$-times heavier than Ni) as well as force constants ($\phi_{\text{Pt-Pt}}$
is $55\%$ larger than $\phi_{\text{Ni-Ni}}$). As such considerable differences
are expected from the lattice dynamical properties of the two systems, with 
different interplay of force constant interactions.
We compare our calculated dynamical matrices, phonon dispersion and DOS
with existing experimental and theoretical data, wherever available.

\begin{table}[t]
\caption{Dynamical matrices $D_{\mu\nu}$($\vert R \vert$) (Newton/meter) for bcc Ta$_{50}$W$_{50}$ (top) and fcc Ni$_{50}$Pt$_{50}$ (bottom). N-atom represents the size of the SQS supercell,  Other experimental\cite{Higuera85} and theoretical\cite{Oscar12} data are give for comparison. }
\centering
\begin{tabular}{ccccccc}\hline\hline
\multicolumn{7}{c}{bcc Ta$_{50}$W$_{50}$} \\ \hline
      & 8-atom & 16-atom & 32-atom & 64-atom & Expt.\cite{Higuera85} &  direction \\ \hline
Ta-Ta &	27.707  & 25.103 & 22.324 & 22.872 & 16.983 & $111_{xx}$ \\
W-W   &	28.934  & 24.568 & 21.063 & 21.069 & 23.000 & $111_{xx}$ \\
Ta-W  &	27.734  & 26.071 & 23.192 & 21.783 & 23.984 & $111_{xx}$ \\
Ta-Ta &	11.504  & 15.727 & 16.120 & 17.599 & 11.201 & $111_{xy}$ \\
W-W   &	 6.655  & 8.348  &  9.418 & 11.156 & 19.200 & $111_{xy}$ \\
Ta-W  &	 8.585  & 13.080 & 13.668 & 13.984 & 17.603 & $111_{xy}$ \\
Ta-Ta &	-0.009  & 9.114  & 25.937 & 17.180 &  1.182 & $200_{xx}$ \\
W-W   &	 0.016  & 17.693 & 36.233 & 41.915 & 47.300 & $200_{xx}$ \\
Ta-W  &	16.448  & 13.442 & 30.733 & 29.367 & 24.803 & $200_{xx}$ \\
Ta-Ta &	-0.001  & 3.105  & -2.909 & -0.598 &  1.423 & $200_{yy}$ \\
W-W   &	 0.008  & 2.755  & -1.648 &  0.244 & -0.800 & $200_{yy}$ \\
Ta-W  &	-1.634  & 3.656  & -2.163 & -0.620 &  1.184 & $200_{yy}$ \\ \hline\hline

\multicolumn{7}{c}{fcc Ni$_{50}$Pt$_{50}$} \\ \hline
      & 8-atom & 16-atom   &  32-atom & 64-atom   & Other\cite{Oscar12} &  direction \\ \hline
Ni-Ni &  6.289  &  8.433 &  9.813 &  9.107 &  8.231 & $110_{xx}$ \\
Pt-Pt & 13.755  & 34.576 & 36.317 &  26.747 &  33.494 & $110_{xx}$ \\
Ni-Pt & 12.421  & 17.387 & 21.210 &  17.200 &  17.868 & $110_{xx}$ \\
Ni-Ni & 3.791  &  8.845 & 11.115 &  9.712  &   9.580 & $110_{xy}$ \\
Pt-Pt &  9.008  & 36.546 & 43.377 &  31.426 &  39.655 & $110_{xy}$ \\
Ni-Pt &  8.026  & 18.287 & 25.091 &  19.416 &  20.740 & $110_{xy}$ \\
Ni-Ni &  5.394  &  0.946 & -1.845 & -0.0423 &  -0.525 & $110_{zz}$ \\
Pt-Pt &  9.487  &  2.310 & -8.351 & -4.199  &  -6.854 & $110_{zz}$ \\
Ni-Pt &  8.785  &  1.035 & -4.124 & -1.141  &  -2.820 & $110_{zz}$ \\ \hline\hline
\end{tabular}
\label{table1}
\end{table}

{\par} To calculate the force constants within a disorder environment,
we first develop a structural model based on the SQS method.\cite{Zunger90}
SQS is an $N$-atom periodic structure constructed in such a way that the
associated set of correlation function of this structure mimic the ensemble 
average correlation functions of the random alloy. Different sized 
SQS-cells ($8$-atom, $16$-atom, $32$-atom and $64$-atom)\cite{SQS-supercell} are used
for both the fcc and bcc systems (See supplementary material [\onlinecite{supplem}] for the SQS structures). We use Vienna ab-initio simulation package
(VASP)\cite{Kresse96} with a pseudo-potential and a projected-augmented-wave
(PAW) basis\cite{Kresse99} based on the local density approximation (LDA). The
cut-off energy for the electronic wavefunctions is $500~e$V. All the structures 
are fully relaxed until the energy converges to within $10^{-6}~e$V and the 
forces on each atom is less than $0.001~e$V/\r{A}. Such a relaxation, in a way, captures the effect of any static displacements that may be present in a real system. A Monkhorst-pack Brillouin 
zone (BZ) integration with a $8^3$ {\bf{k}}-mesh is used for $8$-atom SQS calculation. Convergence of $D_{\mu\nu}$ as a function of {\bf k}-points is checked, see below.
Smaller {\bf{k}}-meshes are used for larger supercells. Magnetic (non-magnetic) calculations are done for NiPt (Ta-W) systems. Relaxed lattice constants for $8$-atom, $16$-atom, $32$-atom and $64$-atom SQS calculation for fcc Ni$_{50}$Pt$_{50}$
are $3.72$, $3.72$, $3.70$, and $3.72$~\r{A} respectively, compared to the
experimental value of $3.785$~\r{A}.\cite{expt_NiPt} For bcc Ta$_{50}$W$_{50}$, 
they are $3.23$~\r{A} for all the structures, compared to $3.23$~\r{A} as 
observed.\cite{expt_TaW} To extract the force constant matrices, we use the
fully relaxed SQS structures and apply the {\it small displacement method}
using PHON package\cite{alfe} implemented within VASP. For 32-atom SQS, force fields are 
constructed by applying $48$ displacements in fcc Ni$_{50}$Pt$_{50}$ and $96$ 
displacements in bcc Ta$_{50}$W$_{50}$ along $3$-cartesian axes, each of 
amplitude $0.04$\r{A}.

{\par} Because of the low symmetry of the SQS-structures, the desired symmetry 
of the dynamical matrices for the underlying lattice (fcc and bcc in the present
case) are lost. Lattice imposed symmetric dynamical matrices are required 
to perform a direct configuration average within the ASR scheme. These 
symmetric matrices are also directly comparable to the neutron scattering data
and other theoretical results, if available. In order to resort to the symmetry
of the underlying lattice, we followed two steps: (1) For a particular SQS-cell
of a given lattice, one can see various atom-pairs along specific crystal
directions followed by the occupancy number for each pair A-A, B-B, and A-B along those directions. This is shown in Fig. S1 of the supplementary material, where the occupancy number for each
pair A-A, B-B, and A-B along various neighbor directions are displayed for 8-, 16- and 32-atoms SQS TaW (left panel) and NiPt (right panel) alloy. Depending on the
occupancy number, the force constants are averaged for each configurations
(i.e. A-A, B-B, and A-B pair) along every directions (up to first nearest neighbor
[nn] for fcc and second neighbor [nnn] for bcc lattices).
(2) Having done the directional averaging, we still lack the proper symmetry of 
the dynamical matrices for the underlying lattice. This is due to the relaxation
effect, which modifies the atomic positions by $\delta$. For example, a 
particular atom at ($1/2$,$1/2$,$0$) in a perfect fcc lattice moves to 
($1/2\pm\delta_{1}^{1}$,$1/2\pm\delta_{2}^{1}$,$\pm\delta_{3}^{1}$) or atoms at
($1/2$,$1/2$,$1/2$) and ($1$,$0$,$0$) in a bcc lattice moves to 
($1/2\pm\delta_{1}^{2}$,$1/2\pm\delta_{2}^{2}$,$1/2\pm\delta_{3}^{2}$) and
($1\pm\delta_{1}^{3}$,$\pm\delta_{2}^{3}$,$\pm\delta_{3}^{3}$) respectively.
In order to retrieve the desired symmetry of the dynamical matrix, we apply 
transformation operation on these average matrices to get the direction
specific dynamical matrices e.g. $\phi_{111} = B_1 \phi_{-111} B_{1}^{T}$ along 
one of the nearest neighbor direction of bcc lattice, where $B_1$ is the 
transformation matrix. A list of transformation matrices along specific 
directions for both fcc and bcc lattices are given in Table S1 of the supplement (See 
[\onlinecite{supplem}]).

{\par} 
The calculated dispersion in the occupancy 
for each pairs reflect the sensitivity of the bond distances on the 
local environment. The
calculated nearest neighbor (next nearest neighbor) average bond distances for 
three pairs d$_{\text{Ta-Ta}}$, d$_{\text{W-W}}$, and d$_{\text{Ta-W}}$ are
$2.837$ ($3.220$), $2.775$ ($3.226$), and $2.788$\r{A} 
($3.235$\r{A}) respectively for $64$-atom SQS Ta$_{50}$W$_{50}$. The nn-bond
distance for Ta-Ta in the alloy is found to be $\sim~0.8\%$ smaller compared
to that in pure Ta, while W-W bond distance in the alloy is $\sim~1.7\%$ 
larger than that in pure W.  The 
calculated dynamical matrices (up to 2nd neighbor) for $8$-, $16$-, $32$- and $64$-atom SQS Ta$_{50}$W$_{50}$ are shown in the upper part of Table \ref{table1}.
Note that, experimental force constants for Ta-Ta 
and W-W pairs are not for the alloy, but for pure Ta\cite{Woods64} and pure
W\cite{Larose76} respectively. Force constants for Ta-W\cite{Higuera85} pair, 
however, are indeed for the alloy. Notably our calculated Ta-Ta force constants
in the alloy are stiffer compared to those in pure Ta. On the other hand, the
calculated W-W force constants are softer  than those in pure W. This prediction
actually jibe with the calculated bond lengths between these two pairs. Alloying
shrinks (expands) the Ta-Ta (W-W) bond lengths making the springs relatively
stiffer (softer). As far as the force constants for Ta-W pair goes, $64$-atom SQS results are our best numbers to compare with the experiment.\cite{Higuera85}
Experimental force constants are computed using a polynomial fit to 
their measured dispersion. Keeping
in mind the sensitivity of the estimates, both on the theoretical and experimental
front, the overall agreement between the $64$-atom SQS results and the experiment
for Ta-W force constants is fairly well.

\begin{figure}[t]
\centering
\includegraphics[width=8.5cm]{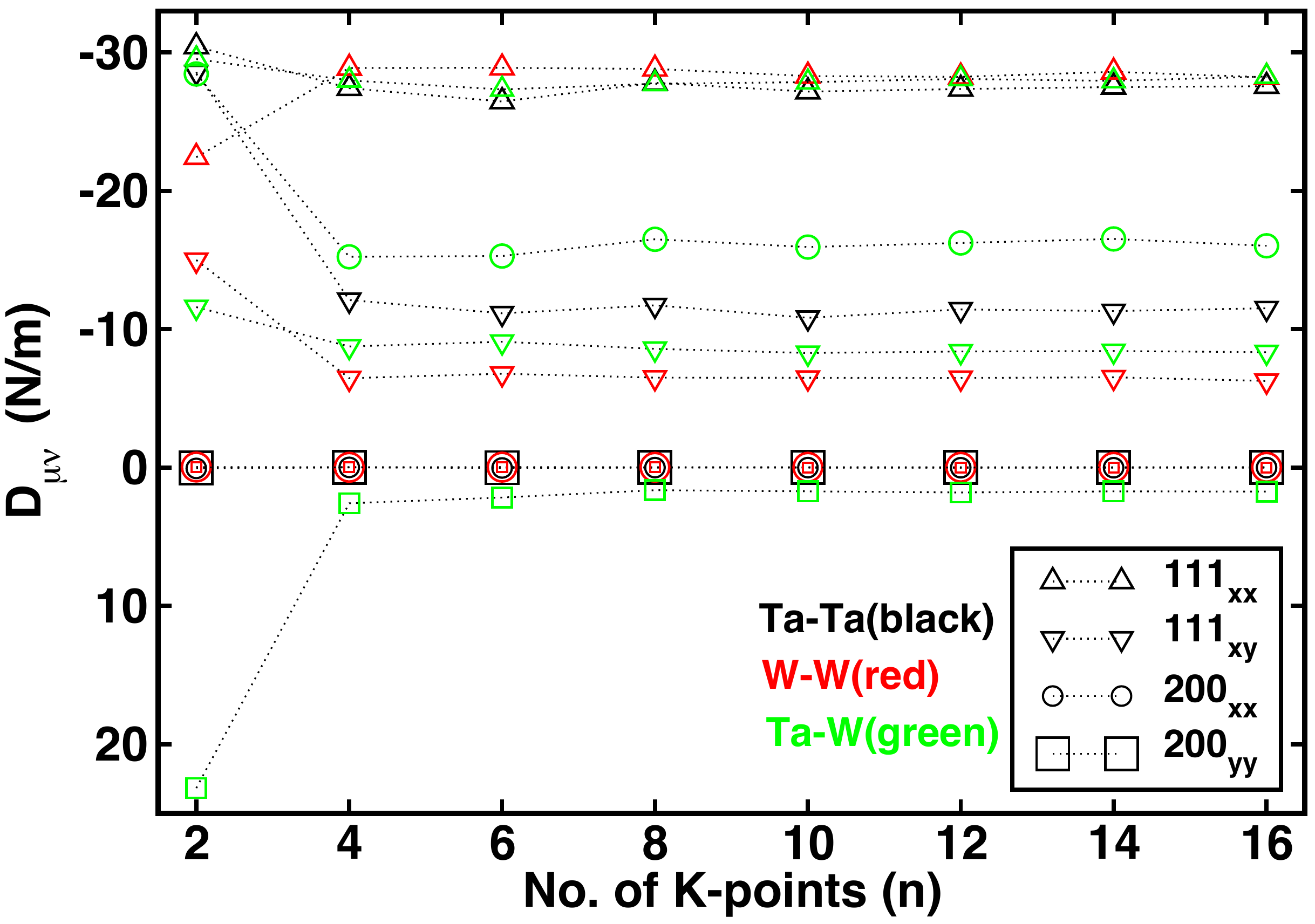}
\caption {(Color online) Convergence of various components of dynamical matrix ($D_{\mu\nu}$) as a function of the number of {\bf k}-points (n, along each direction) for $8$-atom SQS bcc Ta$_{50}$W$_{50}$. Different colors indicate different pairs of force constants, and various components of {\bf D} distinguished by different symbols. }
\label{fig1}
\end{figure}

\begin{figure}[t]
\centering
\includegraphics[width=8.5cm]{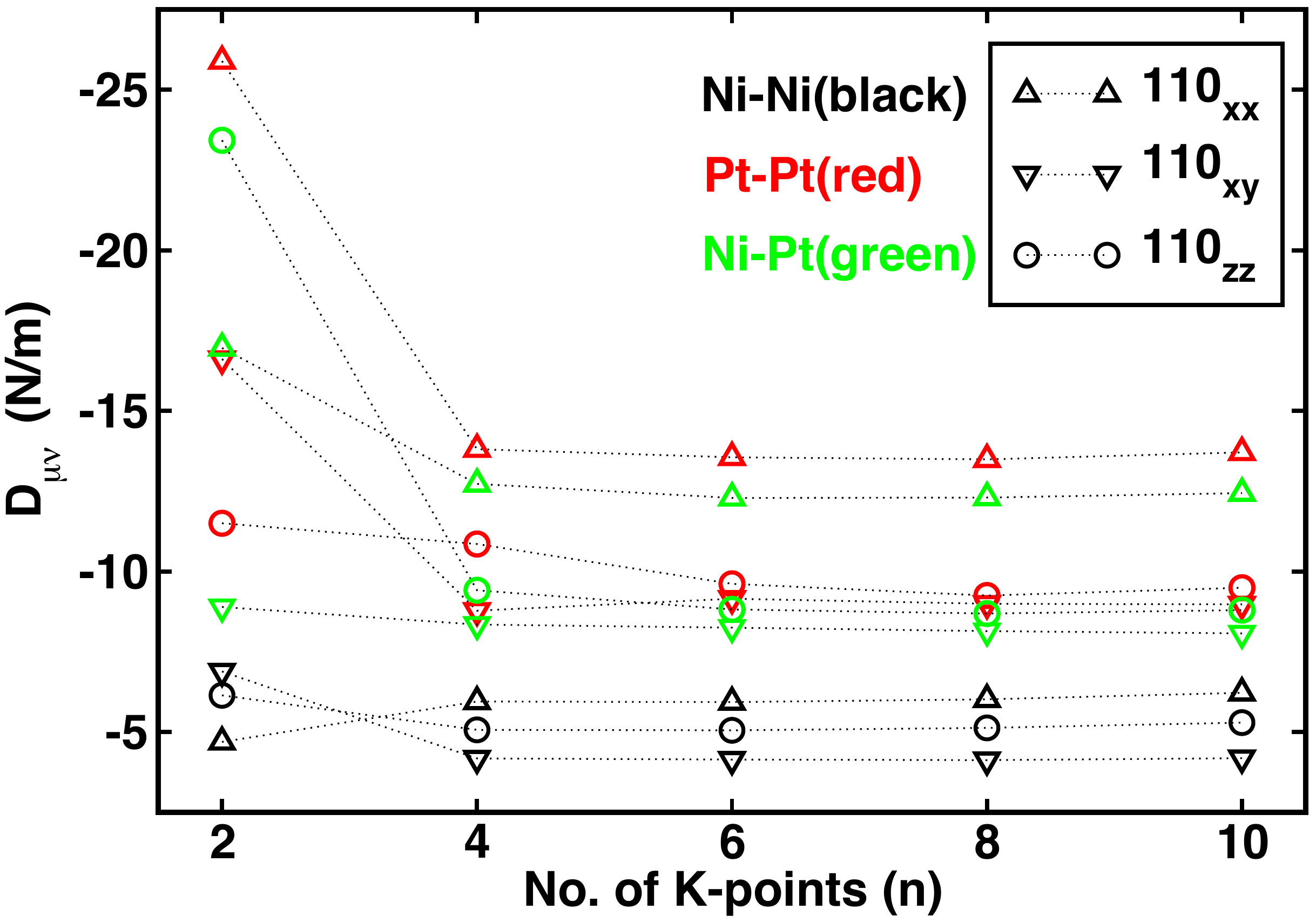}
\caption {(Color online) Same as Fig. \ref{fig1}, but for fcc Ni$_{50}$Pt$_{50}$.}
\label{fig2}
\end{figure}

{\par}
 The calculated average
bond lengths for Ni-Ni, Pt-Pt and Ni-Pt pairs are $2.573$, $2.692$ and
$2.604$\r{A} respectively for $64$-atom SQS Ni$_{50}$Pt$_{50}$. Ni-Ni (Pt-Pt) bond length in the alloy is 
$\sim~3.3\%$ larger ($\sim~2.8\%$ smaller) than those in pure Ni (Pt). As such,
Ni-Ni (Pt-Pt) force constants in the random alloy is expected to get softer
(stiffer) compared to those in pure Ni (Pt). The calculated ab-initio force
constants for the three pairs in disordered Ni$_{50}$Pt$_{50}$ alloy are shown
in the lower panel of Table \ref{table1}. As before, results are shown for 
$8$-, $16$-, $32$- and $64$-atom SQS. Because these calculations are done for 
ferromagnetic NiPt, a one to one comparison with the experiment requires observed data at very low temperature, which we were unable to find in the literature. We have also performed non-magnetic calculations for NiPt, and found similar results for $D_{\mu\nu}$. In Table \ref{table1}, the force constants under the column 
labeled  {\it Other}\cite{Oscar12} are the results from a recent calculation on
Ni$_{50}$Pt$_{50}$ alloy by Granas {\emph {et al.}}\cite{Oscar12} These force
constants for each pair (Ni-Ni, Pt-Pt and Ni-Pt) are within the disordered 
environment, and agree fairly well with ours within a few percent. Calculated
force constants for Ni-Ni (Pt-Pt) pairs in the alloy are found to be softer
(stiffer) compared to those in pure Ni (Pt) (See Ref. \onlinecite{Dutton72} for
the force constants of pure Ni and Pt). This, again, goes in accordance with the
bond lengths of respective pairs in the alloy vs. those in pure elements.

{\par} Keeping in mind the sensitivity of dynamical matrices to finer details of calculation, we have checked the convergence of various components of ${\bf D}$ with respect to the number of {\bf k}-points used in BZ integration. This is shown in Fig. \ref{fig1} and \ref{fig2} for $8$-atom SQS Ta$_{50}$W$_{50}$ and Ni$_{50}$Pt$_{50}$ respectively. Different colors indicate different pairs of force constants while various components of {\bf D} are distinguished by different symbols. One can notice that ALL the components of {\bf D} for both the systems are well converged by $6^3$ {\bf k}-points.  

\begin{figure}[t]
\centering
\includegraphics[width=8.5cm]{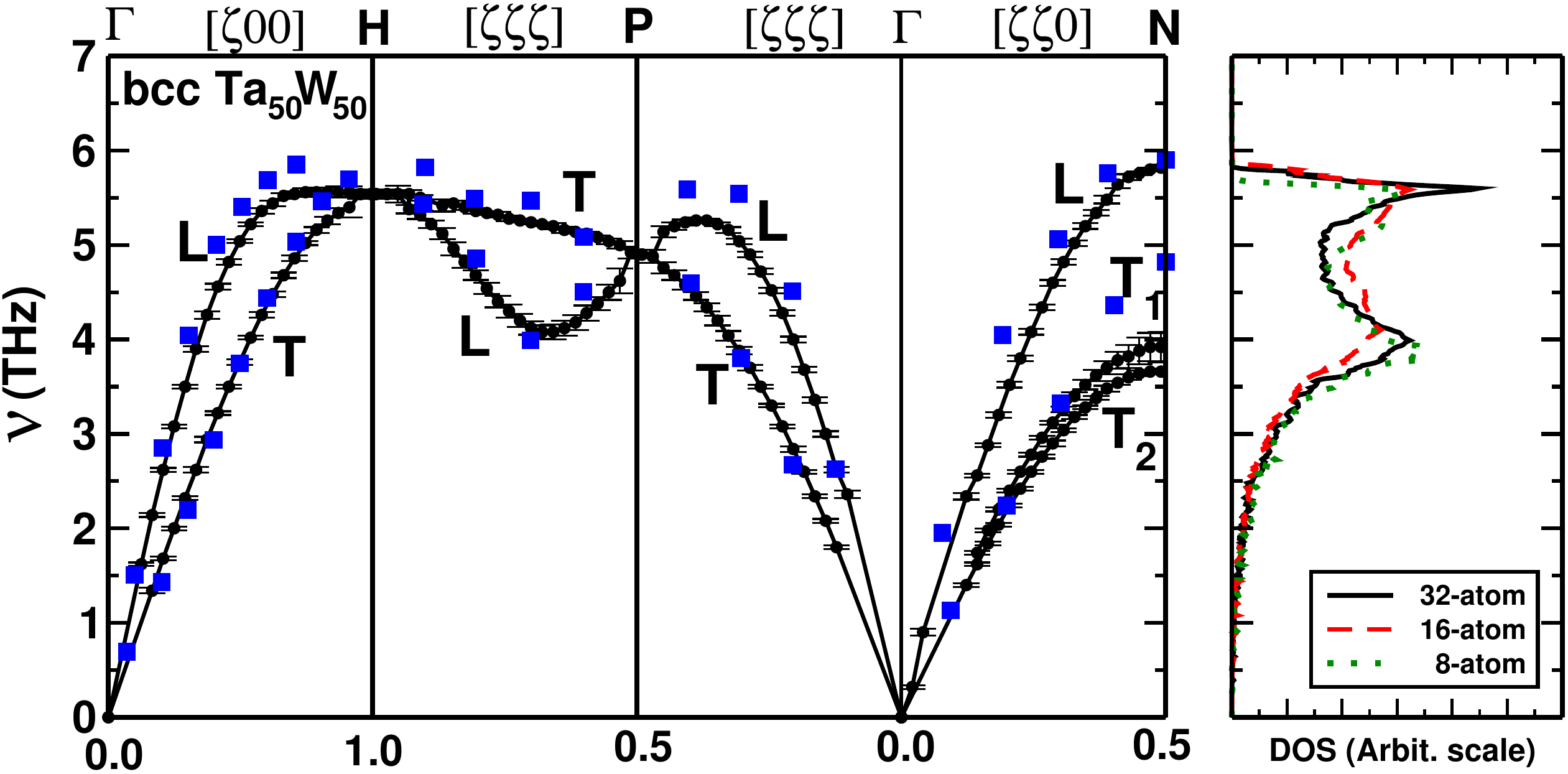}
\caption {(Color online) (Left) Phonon dispersion for bcc Ta$_{50}$W$_{50}$ alloy using the force constants of $32$-atom SQS. L and T stands for longitudinal and transverse modes. Error bars indicate the calculated FWHM's. Square symbols indicate experimental data.\cite{Higuera85} (Right) Phonon DOS using the force constants of $8$-, $16$-, and $32$-atom SQS.}
\label{fig3}
\end{figure}

{\par} Figure \ref{fig3} shows the calculated phonon dispersion (left) and the 
configuration averaged phonon DOS (right) for bcc Ta$_{50}$W$_{50}$. Dispersion
curves are calculated using the force constants of $32$-atom SQS, as listed in 
Table \ref{table1}. Phonon DOS, however, are shown with three sets of
force constants, i.e. $8$-, $16$- and $32$-atom SQS, for comparison.  Error bars
in the dispersion curve indicate the full widths at half maxima (FWHM). Our
calculated phonon dispersion and DOS using $64$-atom SQS force constants are very similar to those using $32$-atom SQS ones (also true in case of NiPt, see below). As such it is intuitively expected that, with increasing size of the supercell, the force constant matrix ${\bf D}$ converges in a collective manner. Finally, note that our calculated dispersion compares fairly well with the experiment
\cite{Higuera85} (shown by square symbol). 

\begin{figure}[t]
\centering
\includegraphics[width=8.5cm]{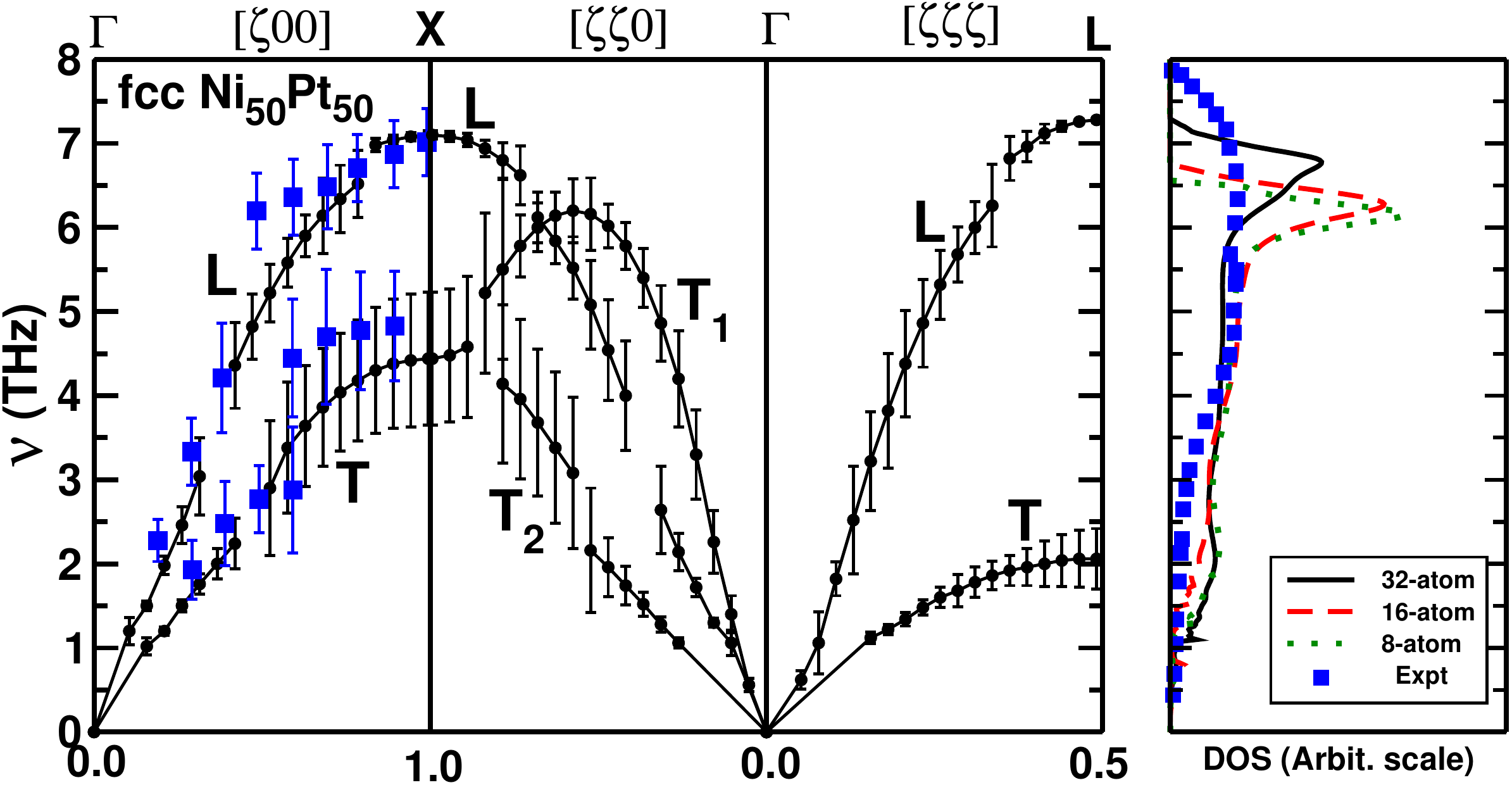}
\caption {(Color online) Same as Fig. \ref{fig3}, but for fcc Ni$_{50}$Pt$_{50}$ alloy. Blue square symbols in both left and right panels indicate the experimental data.\cite{Tsunoda79}}
\label{fig4}
\end{figure}

{\par} Figure \ref{fig4} (left) shows the phonon dispersion for 
Ni$_{50}$Pt$_{50}$ alloy calculated using the force constants of $32$-atom
SQS cell (see Table \ref{table1}). Unlike TaW, NiPt alloy shows interesting 
split band behavior along each symmetry direction. This is due to the strong 
disorder in both mass and force constants, giving rise to resonant modes, 
and has been evidenced in previous studies.\cite{Tsunoda79,Oscar12} Error 
bars with solid circles indicate the calculated FWHM. Error bars with square
symbol along [$\zeta 0 0$] direction indicate the neutron scattering data.
\cite{Tsunoda79} The panel on the right shows the configuration averaged 
phonon DOS with three sets of calculated force constants. Square symbols
indicate the generalized phonon DOS derived from inelastic incoherent scattering.
\cite{Tsunoda79} Notice that the calculated band edge increases with increasing
the SQS cell size, and compare better with experiment. The integral value under
each phonon DOS, however, remain the same. It is important to emphasize that
the experimental phonon DOS is only shown for reference. A one to one
comparison between our calculated DOS and the experimental DOS is not feasible.
In inelastic neutron scattering, phonon DOS can be represented as 
$N(\omega) = \sum_{j} (b_j/M_j) n_j(\omega)$, where $b_j$, $M_j$ and 
$n_j(\omega)$ are the inelastic scattering cross section, atomic mass, and the
partial phonon DOS of atom $j$ respectively. Although the calculated DOS from the
force constants of $32$-atom (or $64$-atom, not shown here) SQS cell resembles maximally with the experimental 
DOS, the calculated band edge is still less than the measured one. This is an 
inherent problem of LDA-based calculations, which usually underestimates the
band edge of the calculated DOS and are also reflected via the bulk modulus.

{\par} In summary, we propose a systematic first principles calculation of the
interatomic force constants for disordered alloys. SQS structures of different 
cell size are used to capture the effects of random environment at different
length scales. Two alloy systems with very different intrinsic properties 
(i.e. lattice type, masses, force constants etc.) are investigated. In bcc
TaW alloy, Ta-Ta force constants are predicted to be stiffer compared to those
in pure Ta, however W-W force constants behave oppositely. In fcc NiPt alloy,
Ni-Ni (Pt-Pt) force constants within the disordered environment behave 
softer (stiffer) than those in pure Ni (Pt). Calculated average bond lengths
between each pair of atoms are found to closely dictate the nature of force 
constants. For both the alloys, the prediction of bond length distribution
and the force constants are found to improve with increasing size of the
SQS-cell; in particular, the $32$-atom and $64$-atom SQS cell for bcc Ta$_{50}$W$_{50}$ yields
force constants which agree fairly well with experiment.\cite{Higuera85}
\textit{Ab-initio} prediction of force constants in random alloys is one of the key
thrust of the present work. Calculated phonon dispersion and the DOS also
compare reasonably well with available experimental data. We propose the 
present method as a potential solution to the microscopic understanding of force
constants in disordered alloys. Future studies on more complex alloys, addressing the effects of magnetic order, lattice mismatch etc. on force constants, are ongoing which will provide an even more stringent test of the present methodology.

We thank D. Alf\`{e} from University college London(UK) for helpful discussion on the symmetry of dynamical matrices.


\end{document}